\documentclass[11pt]{article}
\newtheorem{theorem}{Theorem}[section]

\begingroup
\newtheorem{definition}[theorem]{Definition}

\newtheorem{proposition}[theorem]{Proposition}

\newtheorem{remark}[theorem]{Remark}
\endgroup

\newenvironment{proof} {{\bf Proof.}} {\ }

\begin{document}

\title{Two representation theorems of three-valued structures \\
by means of binary relations
\thanks{Research realized in the framework of COST Action n$^\circ$ 15 (in Informatics)
``Many-Valued Logics for Computer Science Applications''}
}

\author{{\Large Luisa Iturrioz} \\
Universit\'e de Lyon, Claude Bernard Lyon 1 \\
Institut Camille Jordan CNRS UMR 5208 \\
43, boulevard du 11 novembre 1918 \\
69622 Villeurbanne cedex, France 
}

\date{\ }
\maketitle
\thispagestyle{empty}
\noindent{\bf Abstract:}
The results here presented are a continuation of the algebraic research 
line which attempts to find properties of multiple-valued systems based on a poset 
of two agents. 

The aim of this paper is to exhibit two relationships between some three-valued 
structures and binary relations. The established connections are so narrow 
that two representation theorems are obtained.

\


\section{Introduction}

In the domain of reasoning about knowledge, a variety of formalisms have 
been developed for modelling multi-agent co-operation. In the majority of 
cases, the set of involved agents is a nonempty set without any structure, 
the language is a standard modal logic for $n$ agents, and 
the knowledge of an agent is managed as an epistemic operator.
 
\

In order to capture approximation knowledge, an alternative framework to 
model perception of a group $T$ is provided by $n$-valued logic. 
The set of agents is a poset, and the language is based on 
intuitionistic logic. We have in mind to propose a formalism to express properties 
of a poset of two co-operating intelligent agents. We intend here to 
present only algebraic results. 

\

The paper consists of two separate constructions. The first one is 
motivated by the attempt to represent elements in three-valued structures by pairs of 
Boolean elements. The second construction is motivated by the claim given 
in \cite{Dunn82} that representations using relations are more ``natural".

Both constructions are obtained via a Stone-type representation theorem.

\

In \cite{Itu00} we considered a three-valued structure which emerged 
from the formalisation of reasoning with a chain of two agents. 

Throughout this paper we will be concerned with an abstract three-valued structure
related to Moisil ideas \cite{Moi40}, \cite{Moi72}, \cite{BFGR} whose definition is given below. 

Let $(T, \leq)$ be a chain with $T = \{t_1, t_2\}$ and $t_1 \leq t_2$.  
In the applications, $T$ can be considered as a poset of two 
co-operating intelligent agents.

On a distributive lattice $(A, 0, 1,\wedge,\vee)$ with 
zero and unit we are going to define three 
unary operators, noted $C, S_{t_1},S_{t_2}$.  The required properties 
for these operators are the following: 
\begin{itemize}

\item the operators $S_t$, for $t \in \{ t_1,t_2 \}$, are $(0, 1)$-lattice homomorphisms 
from $A$ onto the sublattice $B(A)$ of all complemented elements of $A$ such that $S_t S_w a = S_w a$ 
for all $t, w \in \{ t_1,t_2 \}$;
 
\item $S_{t_1}$ and $S_{t_2}$ are respectively an interior and a closure operator 
on $A$ \\ 
(\cite{Ras74}, pp.$115-116$);
 
\item $S_{t_1}$ is related to the operation C by the equations:\\ 
$S_{t_1} a \wedge Ca = 0$ and $S_{t_1} a \vee Ca = 1$, for all $a \in A$.  
\end{itemize}

This situation suggests the following definition. For notational convenience, 
sometimes we replace $t_1$ and $t_2$ by their indices (i.e., one and 
two).

\section{$T$-structures}
\begin{definition} An abstract algebra $(A, 0, 1,\wedge,\vee,C, S_1,S_2)$
where $0, 1$ are zero-ar\-gument operations, $C,S_1,S_2$ are one 
argument operations and $\wedge,\vee$ are two-ar\-guments operations is 
said to be a {\bf Distributive lattice with three unary operators} if  
\begin{itemize}
\item[](T1) $(A,0,1,\wedge,\vee)$ is a distributive lattice with zero and unit,

\noindent and for every $a,b \in A$ and for all $i,j = 1,2$, the following equations hold:
\item[](T2) $S_i (a \wedge b) =  S_i a \wedge S_i b$ ; $S_i (a \vee b) =  S_i a \vee S_i b,$ 
\item[](T3) $S_1 a \wedge Ca = 0$ ; $S_1 a \vee Ca = 1,$
\item[](T4) $S_i S_j a = S_j a,$
\item[](T5) $S_1 0 = 0$ ; $S_1 1 = 1,$
\item[](T6) If $S_i a = S_i b$, for all $i= 1,2,$ then $a=b,$ (\it Determination Principle)
\item[](T7) $S_1 a \leq S_2 a.$
\end{itemize}
\end{definition}

This definition is not equational.
We will refer to a {\bf $T$-structure} $A$, for short.

\begin{proposition}
The following properties are true in any $T$-structure:
\begin{itemize}
\item[] (T8) $S_2 0 = 0$ ; $S_2 1 = 1,$
\item[] (T9) $a \leq b$ if and only if for $i=1,2$, $S_i a \leq S_i b,$
\item[] (T10) $S_1 a \leq a \leq S_2 a,$
\item[] (T11) $S_i a \wedge CS_i a = 0$ ; $S_i a \vee CS_i a = 1,$ for $i=1,2$.
\end{itemize}
\end{proposition}

\begin{proof}
Indeed by $(T5)$ and $(T4)$ we get $S_2 1 = S_2 S_1 1 = S_1 1 = 1$. 
The proof of $S_2 0 = 0$ is similar. Thus $(T8)$ holds. Assume $a \leq b$, 
i.e. $a = a \wedge b$. By $(T2)$, it follows that $S_i a \leq S_i b$. 
On the other hand, if $S_i a \leq S_i b$, then by $(T2)$, 
$S_i a = S_i a \wedge S_i b = S_i (a \wedge b)$. Hence by the 
determination principle $(T6)$, $a = a \wedge b$ and $a \leq b$. Thus $(T9)$ 
holds. By $(T9)$, the property $(T10)$ is equivalent to 
$S_i S_1 a \leq S_i a \leq S_i S_2 a$, which is equivalent by $(T4)$ to 
$S_1 a \leq S_i a \leq S_2 a$. This together with $(T7)$ proves $(T10)$.
It follows from $(T3)$ that $S_1 S_i a \wedge CS_i a = 0$ and $S_1 S_i a \vee CS_i a = 1$; 
by $(T4)$, $S_i a \wedge CS_i a = 0$ and $S_i a \vee CS_i a = 1$. Thus $CS_i a$ is the 
Boolean complement of $S_i a$, for $i=1, 2$.
\end{proof}

\begin{remark}
Let $\bf B(A)$ be the Boolean algebra of all complemented elements in $A$ 
and $\bf S_i(A)$ the image of $A$ under $S_i$, for all $i=1,2$.   
Since by $(T2)$, $(T5)$ and $(T8)$, mappings $S_i$ are $(0,1)$-lattice 
homomorphisms, $S_i(A)$ is a sublattice of $A$, for all $i=1,2$.

By $(T4)$, $S_i(A) = \{x \in A: S_i x = x\}$ and $\bf S_1(A) = 
S_2(A)$, i.e. mappings $S_i$ have a common image.

By (T11), $S_i(A) \subseteq B(A)$. Using $(T2)$, $(T5)$ and $(T10)$ we get
$\bf S_1(A) = B(A)$. All these proofs can be found in \cite{Itu82}, 
\cite{Itu83}, \cite{ItuVSS}.
 
By (T11), if $``-"$ denotes the Boolean negation we remark that $- S_i a = CS_i a$.
\end{remark}

\begin{proposition}
Let $(A, 0, 1,\wedge,\vee,C, S_1,S_2)$ be a $T$-structure. We define 
two operations $\Rightarrow$ and $\neg$ by means of the following equations, 
for all $a, b \in A:$
\begin{eqnarray}
a \Rightarrow b & = & b \vee \bigwedge_{k=1}^{2} (CS_k a \vee S_k b), \\
\neg a & = & a \Rightarrow 0.
\end{eqnarray}
Then the algebra $(A, 0, 1,\wedge,\vee, \Rightarrow, \neg, S_1, S_2)$ 
is a Heyting algebra with two unary operators satisfying the equation
\begin{eqnarray}
(a \Rightarrow b) \vee (b \Rightarrow a) = 1
\end{eqnarray}
that is, a linearly ordered Heyting algebra \cite{Mont62}, \cite{Mont80}. 
\end{proposition}
\begin{proof}
See \cite{Itu00}, \cite{Itu77}-\cite{Itu83}.
\end{proof}

\

An equivalent {\bf equational} definition of a $T$-structure is given below. 
\begin{definition} A {\bf Heyting algebra with three unary operators} 
(or {\bf $\bf HT$-algebra} for short) 
is an abstract system $A = (A, 0, 1,\wedge,\vee,\Rightarrow,\neg,S_1,S_2)$
such that $0, 1$ are zero-argument operations, $\neg,S_1,S_2$ are one 
argument operations and $\wedge,\vee,\Rightarrow$ are two-arguments 
operations satisfying the following conditions, for all $a, b, c \in A:$  
\begin{itemize}
\item[] (HT1) $(A,0,1,\wedge,\vee,\Rightarrow, \neg)$ is a Heyting algebra,

and for every $a,b \in A$ and for all $i,j = 1,2$ the following equations hold:
\item[] (HT2) $S_i (a \wedge b) =  S_i a \wedge S_i b$ ; $S_i (a \vee b) =  S_i a \vee S_i b,$
\item[] (HT3) $S_2 (a \Rightarrow b) = (S_2 a \Rightarrow S_2 b),$
\item[] (HT4) $S_1 (a \Rightarrow b) = (S_1 a \Rightarrow S_1 b) \wedge (S_2 a \Rightarrow S_2 b),$
\item[] (HT5) $S_i S_j a = S_j a,$
\item[] (HT6) $S_1 a \vee a = a,$
\item[] (HT7) $S_1 a \vee \neg S_1 a = 1, \ \ {\mbox with} \ \  \neg a = a\Rightarrow 0.$
\end{itemize}
\end{definition}

The next two theorems state the equivalence between the notion of 
$T$-structure and that of $HT$-algebra and are proved in \cite{Itu00}.

\begin{theorem}
Let $(A, 0, 1,\wedge,\vee, C, S_1,S_2)$ be a $T$-structure and 
$\Rightarrow$ and $\neg$ be two operations defined by means of the 
following equations, for all $a, b \in A$:
\begin{eqnarray}
a \Rightarrow b & = & b \vee \bigwedge_{k=1}^{2} (CS_k a \vee S_k b), \\
\neg a & = & a \Rightarrow 0.
\end{eqnarray}
Then the algebra $A = (A, 0, 1,\wedge,\vee,\Rightarrow,\neg,S_1,S_2)$ 
is a $HT$-algebra.
\end{theorem}
 
Conversely:
\begin{theorem}
Let $A = (A, 0, 1,\wedge,\vee,\Rightarrow,\neg,S_1,S_2)$ be a 
$HT$-algebra and let us introduce a new operation $C$ by means 
of the following equation, for all $a \in A:$
\begin{eqnarray}
Ca = \neg S_1 a
\end{eqnarray}
Then the abstract algebra $(A, 0, 1,\wedge,\vee, C, S_1,S_2)$ is a $T$-structure.
\end{theorem}

The following general results will be used later on.

\begin{remark} \label{rem:PF}
For a prime filter $M$ in a Heyting algebra, the conditions
\begin{itemize}
\item[] (a) $M$ is maximal among the filters which do not contain the 
element $a$,
\item[] (b) $a \not\in M$ and for every $x \not\in M$, $x \Rightarrow a \in M$ 
\end{itemize}
are equivalent (\cite {Die66}, p.$23$).
\end{remark}

\begin{remark} \label{rem:BMC}
Since $S_2$ and $``\neg\neg"$ are Boolean multiplicative closure 
operators in the sense of \cite{Cig65}, defined on $A$, it follows that
\begin{eqnarray}
S_2 x = \neg\neg x = \bigwedge \{ b \in B(A): x \leq b \}.
\end{eqnarray}
\end{remark}

Two additional facts are recalled for future use.
They concern the prime filters in a $HT$-algebra $A$ and were proved in \cite{Itu00}.

\begin{theorem}
The set of all prime filters in a $HT$-algebra, ordered by inclusion, is 
the disjoint union of chains having one or two elements.
\end{theorem}

\begin{proposition}{\label{prop:P}}
Let $A$ be a $HT$-algebra. If $P$ and $Q$ are two prime filters such that 
$P \subset Q$ and $S_2 x \in P \subset Q$ then $x \in Q$.
\end{proposition}

\section{Examples}
For the sake of illustration let us consider some examples depicting 
the introduced notions. They illustrate our motivations for concrete applications.

\

{\bf 1)} Let  $T = \{t_1, t_2\}$ be an ordered set such that $t_1 \leq t_2$. For each $t \in T$
we denote $F(t)$ the {\bf increasing subset} of $T$ , i.e. 
$$F(t) = \{w \in T : t \leq w \}.$$
Let $A$ be the class of the empty set and all increasing sets, i.e.
$$A = \{ \emptyset, F(t_2), F(t_1)\}.$$ 
The class $A$, ordered by inclusion, is an ordered set with three or two elements, 
and the system $(A, \emptyset, A, \cap, \cup)$, closed under the operations of intersection 
and union, is a distributive lattice with zero and unit. For each $t \in T$ we define 
a special operator $S_t$ on $A$ in the following way:
\begin{eqnarray*}
S_t(F(x)) = & T & \mbox{if} \ \ t \in F(x),\\
S_t(F(x)) = &  \emptyset & otherwise.
\end{eqnarray*}
Finally we define $CF(x) = \neg S_{t_1}(F(x))$. Thus the system $(A, \emptyset, 
T, \cap, \cup,C, S_{t_1},S_{t_2})$ is a $T$-structure, called {\bf basic} 
$T$-structure and denoted $\bf BT$ or $\bf B$ if it has three or two 
elements respectively.

\

{\bf 2)} Let $Ob$ be a nonempty set (set of objects) and $R$ an equivalence 
relation on $Ob$. Let $R^{\ast}$ be the family of all equivalence classes 
of $R$, i.e. $R^{\ast}=\{ \mid x \mid : x \in Ob \}$. This family is a partition 
of $Ob$.
It is well known (see for example \cite{Dav54}, \cite{Mont60}) that on the Boolean algebra 
$B= ({\cal P}(Ob), \emptyset, Ob, \cap, \cup, -)$ where ${\cal P}(Ob)$ denotes the powerset 
of $Ob$, the equivalence relation $R$ induces 
a unary operator $M$ in the following way, for $A \subseteq Ob$:
\[
MA = \bigcup \{\vert x \vert \in R^{\ast} : x \in A\};
\]
which is equivalent to
\[
MA = \bigcup \{\vert x \vert \in R^{\ast} : \vert x \vert \cap A \neq \emptyset\}.
\] 

By definition we have $M(\emptyset)=\emptyset$ and $A \subseteq MA$. 
It is well known (see for example \cite{Dav54}, \cite{Itu99}) that  $M$ also satisfies 
the condition $M(A \cap MB) = MA \cap MB$, for all $A, B \in {\cal P}(Ob)$.

We conclude that $M$ is a monadic operator on the Boolean algebra $B$ 
and that the system $B = ({\cal P}(Ob), \emptyset, Ob, \cap, \cup, - , M)$ 
is a Monadic Boolean algebra \cite{Hal55}, \cite{Hal62}.
As usual we define $LA = -M-A$.

Let $B^{\ast}$ be the collection of pairs $(LA, MA)$, where $A \in {\cal P}(Ob)$. 
Since $LA$ and $MA$ are elements of the Boolean algebra $M({\cal P}(Ob))$ of 
closed elements in $B$ and $LA \subseteq MA$, we consider on $B^{\ast}$ the following
 operations:
\vspace{-0.2 cm}
\begin{eqnarray*}
(LA, MA) \wedge (LB, MB) &=& (LA \cap LB, MA \cap MB)\\
(LA, MA) \vee (LB, MB) &=& (LA \cup LB, MA \cup MB)\\
S_{t_2}(LA, MA) &=& (MA, MA)\\
S_{t_1}(LA, MA) &=& (LA, LA)\\
C(LA, MA) &=& (- LA, - LA)\\
0=(\emptyset,\emptyset) &;& 1=(Ob,Ob)
\end{eqnarray*}

The right side equalities above are in $B^{\ast}$ because the system\\ 
$(M({\cal P}(Ob)), \emptyset, Ob, \cap, \cup, -, M)$ is a monadic 
Boolean subalgebra of $B$.
 
The system $B^{\ast}= (B, 0, 1, \wedge, \vee, C, S_{t_1}, S_{t_2})$ is a 
$T$-structure. By the way of example we check the condition $(T6)$. 
Suppose $S_{t_1}(LA, MA) = S_{t_1}(LB, MB)$ and $S_{t_2}(LA, MA) = 
S_{t_2}(LB, MB)$ then $(LA, LA) = (LB, LB)$ and $(MA, MA) = \\ (MB, MB)$. 
We deduce $LA = LB$ and $MA = MB$ and the pairs $(LA, MA)$ and $(LB, 
MB)$ are equal.

In the literature, a system such as $(Ob, R)$ is called an {\bf approximation 
space} and a pair $(LA, MA)$ is called a {\bf rough set}. They are 
concepts related to {\bf Information systems} in the sense of Pawlak \cite{Paw91}. 

\

{\bf 3)} Let $Ob$  be a nonempty set and let  $g$  be an involution of $Ob$, i.e. a 
mapping from $Ob$  into  $Ob$  such that  $g(g(x)) = x$, for all $x \in Ob$. 
Clearly, every involution $g$ of $Ob$ is a one-one mapping from $Ob$ onto 
$Ob$ and $g = g^{-1}$.
Let us put for each $X \subseteq Ob$ :  
\vspace{-0.2cm}
\begin{eqnarray*}
S_1 X &=& X \cap g(X) \\
C X &=& Ob - (X \cap g(X)) \\ 
S_2 X &=& X \cup g(X). 
\end{eqnarray*}
 
Let $A(Ob)$ be a nonempty class of subsets of  $Ob$, containing 
$\emptyset$ and $Ob$, and closed under set-theoretical intersection and union 
as well as under the operations  $C, S_1$ and  $S_2$  defined above. 
The system $(A(Ob), \emptyset, Ob, \cap, \cup, C, S_1, S_2)$ satisfies the 
conditions:\\
$(T1), (T3), (T4), (T5), (T7)$ and a half of $(T2)$, namely: 
\begin{eqnarray*}
S_1(X \cap Y) &=& S_1 X \cap S_1 Y \\
S_2(X \cup Y) &=& S_2 X \cup S_2 Y.
\end{eqnarray*}

Some subalgebras of $(A(Ob),\emptyset, Ob, \cap,\cup, C, S_1, S_2)$ satisfy 
all the conditions \\ $(T1)-(T7)$, for every $X, Y \subseteq Ob$, i.e. they 
are {\bf $T$-algebras of sets}.
 
These examples are typical, in the sense that every $T$-structure 
is isomorphic to a $T$-structure of sets, as it will be proved in Section 5. 

\

{\bf 4)} If $R$ is a binary relation, we note  $R^{-1} = \{ (y, x) : (x, y) \in 
R \}$ the relation inverse.
Let $E$ be a nonempty set, $\rho$ a fixed {\bf symmetric} relation on $E$  
($\rho \subseteq E \times E$), 
and let  $(A(E, \rho), \emptyset, \rho, \cap, \cup)$  be a lattice of subsets of $\rho$.
 
We can define on  $(A(E, \rho), \emptyset, \rho, \cap, \cup)$  the operations $S_1$, 
$S_2$ and $C$ in the following way, for  $R \subseteq \rho~:$
\vspace{-0.2cm}
\begin{eqnarray*}
S_1(R) &=& R \cap R^{-1} \\
S_2(R) &=& R \cup R^{-1} \\
C(R) &=& \rho - S_1(R).
\end{eqnarray*}

The system $(A(E, \rho), \emptyset, \rho, \cap, \cup, C, S_1, S_2)$ 
satisfies the conditions $(T1)$, $(T3)$, \\ $(T4)$, $(T5)$, $(T7)$ and a half of 
$(T2)$, as in example 3.

Some subalgebras of $(A(E, \rho), \emptyset, \rho, \cap, \cup, C, S_1, S_2)$ satisfy 
all the conditions $(T1)-(T7)$, for every $X, Y \subseteq E$, i.e. they 
are {\bf $T$-algebras of relations}.

\section{First construction}

In this section we recall the proof of a theorem given in \cite{Itu00}, which exhibits a 
method to construct a concrete $T$-structure.

\

Let  $A$  be a $HT$-algebra. By Theorem 2.9, the set $Ob$ of all prime filters in $A$, 
ordered by inclusion, is the disjoint union of chains having one or two elements. 

\bigskip

Let $R_{Ob}$ be the {\bf binary relation} defined on $Ob$ in the following way:

\bigskip

If $P, Q \in Ob$ then we put $P R_{Ob} Q$ if and only if $P$ and $Q$ 
are comparable,\\ 
i.e. if they are in the same chain. $R_{Ob}$ is an 
{\bf equivalence relation} on $Ob$.

\bigskip

We consider the Monadic Boolean algebra 
$({\cal P}(Ob), \emptyset, Ob, \cap, \cup, -, M)$, where for $X \subseteq Ob:$
\[
MX = \bigcup \{\vert P \vert \in R^{\ast}_{Ob} : P \in X\}.
\]

Following Stone, for every $x \in A$ we define the map 
$s: A \rightarrow {\cal P}(Ob)$ as follows:
$s(x) = \{ P \in Ob : x \in P \}$. The map $s$ is a one-one 
$(0,1)$-lattice homomorphism.

\bigskip

Let $B^{\ast}$ be the collection of pairs $(Ls(x), Ms(x))$ with 
operations defined as in example 2. The system $(B^{\ast}, 
\emptyset, Ob, \cap, \cup, C, S_1, S_2)$ is a $T$-structure.
We consider the map $h: A \rightarrow B^{\ast}$ defined as 
follows: $h(x) = (Ls(x), Ms(x))$.

\bigskip

This leads to the result below, showed in \cite{Itu00}.

\begin{theorem}{\bf Representation theorem}. Every $HT$-algebra can 
be represented as an algebra of rough subsets of an approximation space $(Ob, R)$.
\end{theorem}

\section{Second construction}

Let $A$ be a $HT$-algebra and let  $E$  be the set of all prime filters in $A$, 
ordered by inclusion. According to Theorem 2.9, the ordered set $(E, 
\subseteq)$ is a disjoint union of chains having one or two elements.
 
We define the map $g : E \rightarrow E$ in the following way:
\vspace{-0.2cm}
\[
g(P) = \left \{
\begin{array}{llll}
P \ , & \mbox {if P is maximal and minimal at the same time,}\\
Q \ , & \mbox {if P and Q are in the same chain and $P \neq  Q$.}
\end{array}
\right. 
\]

\

The map $g$ is an involution of $E$. For each $X \subseteq E$ we define the 
operations $S_1$, $C$ and $S_2$ as in example 3.
Let  $f : A \rightarrow {\cal P} (E)$  be the Stone isomorphism,  i.e. 
for each  $a \in A$, 
$f(a) = \{P \in E : a \in P \}$. It is well know that $f$ is a 
one-one $(0,1)$-lattice homomorphism. 

We show that $f$ satisfies also the conditions:
\[
f(S_1 a) = S_1 f(a),\ \  f(S_2 a) = S_2 f(a),\ \ f(Ca) = Cf(a).
\]
By the way of example we show the condition $f(S_2 a) = S_2 f(a)$. The 
proof of this equality is accomplished in four steps:
\vspace{-0.2cm}
\begin{itemize}
\item[(i)] $f(S_2 a) \subseteq S_2(f(S_2 a))$. Immediate from the 
definition of $S_2$.
\item[(ii)] $S_2(f(S_2 a)) \subseteq f(S_2 a)$. Assume $P \in S_2(f(S_2 a)) = 
f(S_2 a) \cup g(f(S_2 a))$. If $P \in f(S_2 a)$ then $S_2 a \in P$ 
and the result is true. If $P \in g(f(S_2 a))$ then $S_2 a \in g(P)$; if $S_2a 
\not \in P$ then $P \subset g(P)$. In this case, $\neg S_2 a \in P$ and 
$S_2 a \wedge \neg S_2 a = 0 \in g(P)$, a contradiction.
\item[(iii)] $f(S_2 a) \subseteq S_2 f(a)$. Let $P \in f(S_2 a)$, 
i.e. $S_2 a \in P$. 
We show that $a \in P$ or $a \in g(P)$. We distinguish three cases. 
Assume $P$ is minimal and $P \subset g(P)$; by Proposition \ref{prop:P} we have $a \in g(P)$.
Assume $g(P) \subset P$. Since $S_2 a \vee \neg S_2 a = 1 \in g(P)$ we 
deduce either $S_2 a \in g(P)$ or $\neg S_2 a \in g(P)$. If $\neg S_2 a \in 
g(P)$ we would have $\neg S_2 a \wedge S_2 a = 0 \in P$ which is 
impossible, so $S_2 a \in g(P)$. By Proposition \ref{prop:P} again, we get $a \in P$.
Assume $P$ is minimal and maximal, then $P = g(P)$. If $a \not \in P$ then 
by Remark \ref{rem:PF} we have $\neg a \in P$, and by Remark \ref{rem:BMC} 
it follows that $\neg a \wedge S_2 a = \neg a \wedge \neg \neg a = 0 \in P$ a contradiction.
We have shown that either $P \in f(a)$ or $g(P) \in f(a)$.
In both cases we conclude $P \in f(a) \cup g(f(a)) = S_2 f(a)$.
\item[(iv)] $S_2 f(a) \subseteq f(S_2 a)$. Since $a \leq S_2 a$ then 
$f(a) \subseteq f(S_2 a)$ and $S_2 f(a) \subseteq S_2 f(S_2 a) = f(S_2 a)$ 
by (i) and (ii) above. 
\end{itemize}

The image $f(A)$ is a {\bf $T$-algebra of sets}.

The set  $G = \{ (P , g(P)) \}_{P \in E}$  is a {\bf 
symmetric relation} on $E$.

\

We consider the map $h: A \rightarrow G \cap (\{ f(a) \}_{a \in A} \times E)$ 
defined by:  
\[
h(a) = G \cap (f(a) \times E).
\]
 
This map $h$ preserves all the operations on $A$. In fact :
\vspace{-0.2cm}
\begin{enumerate}
\item $h$ is one-one. 

In fact, if  $a \neq b$ then  $f(a) \neq f(b)$ (Stone). 
Suppose without loss of generality that  $x  \in f(a)$  and  $x \not \in f(b)$  
then $(x, g(x)) \in G \cap (f(a) \times E)$ , but  
$(x, g(x)) \not \in G \cap (f(b) \times E)$; thus $h(a) \neq h(b)$ as desired.
\item $h(a \wedge b) = h(a) \cap h(b)$, \ $h(a \vee b) = h(a) \cup h(b)$.

$h(a \wedge b) = G \cap (f(a \wedge b) \times E) = G \cap (f(a) \cap f(b)) \times E) =\\ 
G \cap [f(a) \times E) \cap (f(b) \times E)] = 
G \cap (f(a) \times E) \cap G \cap (f(b) \times E) = h(a) \cap h(b)$.\\
The proof of the other equality is similar.
\item $h(S_2 a) = S_2 h(a)$, \ $h(S_1 a) = S_1 h(a)$, \ $h(C a) = C h(a)$.

$h(S_2 a) = G \cap  ( f(S_2 a) \times E) = 
G \cap (S_2 f(a) \times E) =  G \cap ((f(a) \cup g(f(a))) \times E) =
[ G \cap (f(a) \times E)] \cup [G \cap (g(f(a)) \times E)].$ On the other 
hand,\\
$S_2 h(a) = h(a) \cup (h(a))^{-1} = 
[G \cap (f(a) \times E)] \cup [G \cap  (f(a) \times E )]^{-1} =$ \\
$[G \cap (f(a) \times E)] \cup [G^{-1} \cap (f(a) \times E)^{-1}] =
[G \cap (f(a) \times E)] \cup [G \cap (E \times f(a))].$ \\ 
We show that  $[G \cap (g(f(a)) \times E)] = [G \cap ( E \times f(a))]$. In fact, it is a 
consequence of the following equivalent conditions: \\
$(x, y) \in  G \cap (g(f(a)) \times E) \Leftrightarrow (x, y) \in G, y = g(x)$ 
and $x \in g(f(a))$ 
$\Leftrightarrow (x, y) \in G, g(x) = y \in f(a)  \Leftrightarrow (x, y) \in 
G \cap (E \times f(a))$.\\
The proof of the other two equalities are similar.
\end{enumerate}

\begin{remark}
The operation $S_2$ defined above, satisfies the following inequalities,\\  
for $R, S \subseteq \rho$ \ :
\vspace{-0.2cm}
\begin{eqnarray*}
S_2(R \cap S) & \subseteq & S_2R \cap S_2S,\\
S_1(R \cup S) & \supseteq & S_1R \cup S_1S.
\end{eqnarray*}
\end{remark}

In fact, $S_2(R \cap S) = (R \cap S) \cup (R \cap S)^{-1} = 
(R \cap S) \cup (R^{-1} \cap S^{-1})$  and
$S_2(R) \cap S_2(S) = (R \cup R^{-1}) \cap (S \cup S^{-1}) = 
(R \cap S) \cup (R \cap S^{-1}) \cup (R^{-1} \cap S) \cup (R^{-1} \cap S^{-1})$.

In the other case the proof is similar.

\

In general, the equalities are not true, i.e. the system is not a 
$T$-structure. Nevertheless, some subalgebras of this system may be.
We close the paper with the following result.

\

We claim that the 
$h$-image $(G \cap (\{ f(a) \}_{a \in A} \times E), \emptyset, G, \cap, \cup, C,
S_1, S_2)$ of  $A$  is a $T$-structure of relations isomorphic to $A$.

By the way of example we show one of the conditions in $(T2)$: 
\vspace{-0.2cm}  
$$
S_2 (h(a) \cap h(b)) =  S_2 (h(a)) \cap S_2(h(b)).
$$
In fact, taking into account the condition $(T2)$ in $A$ and 
the fact that $h$ is a homomorphism, we get:

\vspace{0.1cm}
$S_2 (h(a) \cap h(b)) =  S_2 (h(a \wedge b)) = 
h(S_2 (a \wedge b)) = h(S_2 (a) \wedge S_2 (b)) =\\
h(S_2 (a)) \cap h(S_2 (b)) = S_2 (h(a)) \cap S_2 (h(b))$. 

\

This completes the proof of the following statement.

\begin{theorem}{\bf Representation theorem}.

Every $T$-structure $(A, 0, 1, \wedge, \vee, C, S_1, S_2)$ 
is isomorphic to a $T$-structure of relations.
\end{theorem}

\noindent {\bf Aknowledgements}. 

The author wishes to thank the two referees of the 31st IEEE International Symposium on Multiple-Valued Logic (ISMVL'2001) for helpful constructive comments. Academic duties prevented the author to attempt the Symposium.

\end{document}